# The Evolution of Raw Data Archiving and the Growth of Its Importance in Crystallography

John R. Helliwell[1], James R. Hester[2], Loes Kroon-Batenburg[3], Brian McMahon[4] and Selina L. S. Storm[5*]

[1] Department of Chemistry, University of Manchester, Manchester M13 9PL, United Kingdom

[2] Australian Nuclear Science and Technology Organisation, Locked Bag 2001, Kirrawee DC 2232 Australia

[3] Structural Biochemistry, Bijvoet Center for Biomolecular Research, Utrecht University, Universiteitsweg 99, 3584 CG Utrecht, The Netherlands

[4] International Union of Crystallography, 5 Abbey Square, Chester CH1 2HU, United Kingdom

[5] European Molecular Biology Laboratory (EMBL) Hamburg, c/o DESY, Notkestraße 85, 22607, Hamburg, Germany

*Correspondence e-mail: selina.storm@embl-hamburg.de

**Synopsis**

The IUCr 75th Congress in Melbourne included a workshop on raw data reuse. This built on earlier IUCr funded workshops on raw data archiving potential in the past decade for establishing the ground truth of research results. This article charts the efforts of IUCr to facilitate discussions and plans within the various communities of crystallography, diffraction and scattering of raw data archiving and reuse.

**Abstract**

The hardware for data archiving has expanded capacities for digital storage enormously in the past decade or more. The IUCr evaluated the costs and benefits of this within an official Working Group which advised that raw data archiving would allow ground truth reproducibility in published studies. Consultations of the IUCr's Commissions ensued via a newly constituted standing advisory committee, the Committee on Data. At all stages the IUCr financed workshops to facilitate community discussions and possible ways of implementation of raw data archiving. The recent launch of *IUCrData* journal's *Raw Data Letters* is a milestone in the implementation of raw data archiving beyond the currently published studies: it includes those diffraction patterns that have not been fully interpreted if at all. The IUCr 75th Congress in Melbourne included a workshop on raw data reuse discussing the successes and ongoing challenges of raw data reuse. This article charts the efforts of IUCr to facilitate discussions and plans relating to raw data archiving and reuse within the various communities of crystallography, diffraction, and scattering.

1. Introduction

A major effort of the IUCr Diffraction Data Deposition Working Group (DDDWG, 2011 to 2017) and now the IUCr Committee on Data (CommDat, since 2017) has been exploring the practicalities, the costs and benefits, and the opportunities for new crystallographic science arising from large-capacity data archives that have become available. We held a full-day workshop at IUCr2023 entitled "Raw diffraction data reuse: the good, the bad and the challenging", bringing to a focus twelve years of work in which we discussed: (1) current practices in raw data archival and sharing; (2) educating those who generate and deal in crystallographic data on best practices in data reuse in various categories of crystallographic science, with talks by leading experts; and (3) we offered a summing up, including the role of *IUCrData*'s new section *Raw Data Letters*. Attendees learnt about the opportunities for raw data reuse, including the use of raw data as test data sets for machine learning, and achieving an understanding of how, and where, to effectively archive their own raw data to maximise the potential for data sharing and reuse in the future.

This workshop explored in detail the successes and challenges in practice of raw data sharing and reuse. Being a full day, it complemented the main Congress microsymposium entitled "Raw diffraction data reuse: warts and all". A second microsymposium of the

Committee on Data as principal proposer was entitled "Interoperability of Data and Databases" (Brink et al 2024). We also secured a Congress keynote presented by Andy Götz of ESRF on the "European Photon and Neutron Open Science Cloud", this being the world leading effort of this consortium of more than ten European synchrotron, neutron and X-ray laser radiation sources with raw data management and sharing.

This article has several roles. Firstly, it provides an overview of the topics addressed in the past twelve years by the IUCr's DDDWG and Committee on Data. Secondly, it sets the scene of the international landscape on raw, processed, and derived data, ensuring reproducibility of science as a whole, and thereby informs our own efforts for the best reproducibility of published crystallographic science. Thirdly, it serves as an introduction to the whole virtual issue of articles from the speakers and poster presenters at IUCr2023. Fourthly, it highlights the important role of standards in the peer review of raw diffraction data, notably via enabling automated tools, which are important for ensuring standards for a Raw Data Letter within *IUCrData*. Also, tools adopted for peer review could also ensure raw data quality at the measuring instruments such as synchrotron crystallography beamlines themselves. This latter, we hope, will also assist in increasing the fraction of published studies from measured datasets. Finally, in a 'future vision' section we note that raw data flows continue to increase substantially with improved sources and detectors along with tackling ever more challenging experiments. There is a balance to be struck therefore between compellingly good principles, such as reproducibility of published work, and the need to be pragmatic in terms of what unpublished raw data are preserved and for how long.

2. Truth and objectivity in crystallographic science and the role of peer review

In a scientific investigation, with one method alone, such as crystallography, we can seek as precise a result as possible, but it will inevitably, to a greater or lesser degree, be inaccurate. That degree depends on controlling as much as possible systematic errors in that one method. By using one or more complementary methods, each with a different set of systematic errors, we learn how well the several methods' results agree, and gain an insight into the accuracy overall. In molecular structure science an example of a popular combinations of methods would be crystal structure analysis and NMR (in solution or the solid state). This theme of combining methods is explored in detail in Helliwell & Massera

(2022) within two subject areas of structural molecular science, respectively chemical and biological; the authors examine reproducibility, replicability, reliability and reusability in defining trust in a scientific study.

On submission of a publication the assessment of the validity of its results, and usually also the study's significance, is undertaken as either pre-publication or post-publication peer review, usually both. Traditionally the pre-publication peer review procedure involves an editor, usually known to the authors, who then consults two or more referees. Usually, the authors are not anonymous to those referees. The editor takes a final decision. These reports are usually not published. Traditionally the post-publication peer review procedure is that a publication has a readership and individual laboratories may go so far as to first check its reproducibility and then may be inspired to design their own study to replicate and or extend the discoveries reported in that publication. Where there are concerns these can be described in a critique article and the original authors are invited to respond. Peer review procedures are evolving away from this traditional model, though. Some journals, at an article's publication, also publish the peer review reports, responses of authors and even the editor's decision letter. Post-publication peer review can now also take the form of the article being published immediately and the readership at large having the opportunity to post comments at the journal's website. Preprint servers have a wide role in involving the community at large as other disciplines have adopted the lead of arXiv, set up by physicists. The challenge for preprint servers is now to include the underpinning data and metadata and, where available, machine-generated consistency checks such as *checkCIF* or PDB validation reports in order to give the reader of a preprint a complete view of the provenance of a study.

In experimental science the effectiveness of peer review assessment is maximised by going back as far as possible in the data records underpinning those results. To borrow a term from machine learning the raw data form the *ground truth* whereas the subsequent processed data will have involved subjective choices made by the researcher both in choosing a particular software approach, and then within the software chosen. Then the derived model fit to those processed data will have involved further sets of subjective choices, again in choosing one software or another and then within the software chosen. Going back as far as possible in the recorded workflow, and data files, takes us as close as possible to objectivity itself. Assessment of any study benefits from as complete as possible a workflow record and its reproducibility assessment by a referee as distinct from a member of the primary research team. The trend of digitally recording all steps of the experiments in electronic logbooks is an important one in this context (https://www.daphne4nfdi.de/TA1.php). There are other aspects besides the digital ones of course, such as the choice of apparatus, beamline or detector and the stability of their calibrations, as well as the choice of sample itself, which should also be documented.

Crystallographers are one community of several (such as astronomers and particle physicists) that have exploited digital data storage media to archive as much as possible of their data. This started with atomic coordinates; then the processed diffraction data (usually the structure factors) were added as the archiving capacity expanded. It is only in the past 15 years or so that it became practical to archive the raw data. The scale of archived data sets is typically kilobytes for a file of coordinates (including their atomic displacement parameters), to megabytes for the structure factors file, to gigabytes for the zipped diffraction images file per crystal structure.

In the years since the formation of the IUCr's Diffraction Data Deposition Working Group in 2011 and its final report delivered in 2017 (https://www.iucr.org/resources/data/dddwg/final-report), two crystal structure communities have engaged carefully, via their Commissions, in the question of the value of archiving raw diffraction images. These are the Biological Macromolecules and Structural Chemistry communities. The former have firmly recommended the archiving of diffraction images for any publication communicating a new structure or a new method (for the implementation in IUCr Journals see Helliwell *et al.*, 2019). By contrast the structural chemistry community, via consultations thus far involving a questionnaire and a workshop, have reported that the procedures for raw diffraction data processing are so satisfactory that only in extreme examples of samples with unusually challenging diffraction need their diffraction images be archived. In effect the chemical crystallographers view their *processed* data, the structure factors, as their ground truth. Despite these broad consensuses, there are others in the structural chemistry community who have advocated keeping raw diffraction images nevertheless, such as for a wider application of quantum chemistry analyses (formerly known as charge density studies). In the structural biology community, there are those, not as enthusiastic about archiving diffraction images as the Commission on Biological Macromolecules, who advocate going only so far back in the data processing workflow that the unmerged structure factors are retained. This latter does allow a better diagnostic of any given experimental data set (than merged processed data to a unique set), and its measurement timeline in particular, such as diagnosing X-ray radiation damage to the sample. It does not avoid the various subjective choices made within a raw data processing software nor the choice of which software out of several available.

Another significant community is that of powder diffractionists. The International Center for Diffraction Data (ICDD; https://www.icdd.com/) have approximately half a million entries overall and include ~10,000 raw diffraction profiles (i.e. without the background stripped off, rather than 2D detector diffraction images). Aranda (2018) has offered views on benefits and challenges of sharing powder diffraction raw data. The Commission on Powder Diffraction is considering the possibilities in detail. Likewise the small angle scattering community has advanced, agreed, protocols for the management and sharing of their measured scattering data (Trewhella et al 2017).

Overall, the DDDWG recommendations can secure the best practical reproducibility of a structure derived from diffraction methods and are consistent with recommendations such as the recent USA report on best practice for *Reproducibility and Replicability in Science* (National Academies of Sciences, Engineering, and Medicine, 2019).

This report, and our narrative above, does not consider unpublished data. To maximise the benefit of funds invested in science, and its facilities, scientific practice surely can and must involve maximising the number of communicated results. This is a different issue from the reproducibility of a publication. Furthermore, measured data flows are accelerating considerably such as at the new extremely bright synchrotrons and the high-data-rate electronic area detectors on beamlines (see *e.g.* Leonarski, 2023). It is now the case that the feasibility of preserving all data in every beamtime shift is under challenge. As Leonarski (2023) neatly puts it *"X-ray facilities may have to make a difficult decision: either expand investment budget in IT infrastructure dramatically or restrict experiment performance"*. Yet of course the fraction that does lead to publication is hard to predict, let alone how long to archive the unpublished raw data before taking the decision to delete. As facilities increasingly move into structural dynamics studies, quantification of the precision and accuracy of atom movements in crystal structures, biological and chemical, will need to be ever more rigorous, not less. Since many data sets in serial crystallography contain empty frames there is no value of storing those frames. Facility data policies such as at the EuroXFEL are being revised (Dell Antonia et al EuroXFEL Data Analysis Workshop 2024 and https://www.xfel.eu/sites/sites_custom/site_xfel/content/e51499/e141242/e141245/xfel_file234453/ScientificDataPolicy2023_eng.pdf ).

3. FAIR and FACT across the sciences; CODATA, Research Data Alliance and World Data System initiatives

The wider science community is increasingly embracing the FAIR data paradigm, namely, that data be Findable, Accessible, Interoperable, and Reusable (Wilkinson *et al.*, 2016). Some social scientists also emphasize that more than FAIR is needed. Their data should be "FACT," which is an acronym referring to Fairness, Accuracy, Confidentiality, and Transparency (van der Aalst *et al*., 2017). These qualities are essential to ensure reproducibility not just reusability. This viewpoint of FAIR and FACT is similar to the view of IUCr as to the importance of data quality and completeness (Hackert et al 2016); see below.

While FAIR looks at practical issues related to the sharing and distribution of data, FACT focuses more on the foundational scientific challenges. Although van der Aalst *et al*. (2017) are writing from the perspective of the social, rather than physical or biomedical sciences, there are aspects of their recommendations that apply to all the sciences. In particular, the requirements for accuracy and transparency emphasise the need to work towards the highest quality in experimental data. The recent (2018) merger of the International Council for Science with the International Social Sciences Council (to form the International Science Council, ISC) is a welcome move towards encouraging a common level of rigour across both the social and the physical/biomedical sciences.

 In crystallography, the requirement for FAIR data is satisfied by our databases for processed diffraction data and their derived molecular models.

However, the FAIR data principles do not contain an explicit reference to the quality of data. The omission of this criterion by Wilkinson *et al.* (2016) may be traced to an influential OECD report of 2007; quoting from that report's section on Quality (OECD, 2007):

> *"The value and utility of research data depends, to a large extent, on the quality of the data itself. Data managers, and data collection organisations, should pay particular attention to ensuring compliance with explicit quality standards. … Although all areas of research can benefit from improved data quality, some require much more stringent standards than others. For this reason alone, universal data quality standards are not practical."*

The same report also states '*Where such standards do not yet exist, institutions and research associations should engage with their research community on their development.*' The IUCr communities have always taken data quality seriously. As an example of a clear, overarching statement for all our crystallographic communities, the IUCr published a considered response (Hackert *et al*., 2016) to the ISC's report *Open Data in a Big Data World* (ISC, 2015). Within this Hackert *et al*. (2016) noted the importance of the publication of this international Accord on the values of open data in the emerging scientific culture of big data and endorsed its analysis of the value of open data. They also emphasised the generality of the Accord, and went on to emphasise the crucial importance of quality control, informed by the practice within crystallography and related structural sciences:

> *"All scientific data must be subject to rigorous first analysis to exclude or quantify systematic bias or error; all software implementations should employ open algorithmic procedures and their results should ideally be cross-checked by independent implementations. An overlooked challenge in handling ever-growing volumes of data is the need to apply the same level of critical evaluation as has been applied to historically smaller volumes….*
>
> *We hold that the essential component of openness is that the data supporting any scientific assertion should be*

- ***complete*** *(i.e. all data collected for a particular purpose should be available for subsequent re-use); and*
- ***precise*** *(the meaning of each datum is fully defined, processing parameters are fully specified and quantified, statistical uncertainties evaluated and declared).*

*Together, these properties include the criteria … that open data should be discoverable, accessible, intelligible, assessable and usable. We note, however, that a full understanding of the data may depend on associated scientific publications that discuss the details of data processing where these differ from routine practice. The full linking of article and data is another key element of openness."*

Since the publication of the FAIR principles by Wilkinson *et al.* (2016) other communities affiliated to CODATA (the ISC Committee on Data, where such matters get debated) have been pushing for a revisit of the omission of data quality in the FAIR principles. This is driven largely by the, obviously compelling, efforts of CODATA to ensure cross-domain integration of interdisciplinary data in tackling challenges such as disaster risk reduction. As an example, crystallography contributed to understanding the covid 19 pandemic through its covid 19 viral protein structures. These aspects of interoperability are discussed in more detail by Brink *et al*. (2024) in a companion article in this IUCr 75$^{th}$ Committee on Data special issue.

Let us consider how elements of the FACT criteria could be applied within the crystallographic sciences, beginning with the notion of 'fairness'. Consider, for example, results from neutron crystallography. Because neutrons are a non-damaging probe of the structure of matter these measurements can be made under ambient conditions, even the conditions of a living cell in terms of, say, temperature and pressure. Hence neutron crystal structures can be regarded in this sense as the closest to truth that we can reach with our atomic-scale probes of the structure of matter, that also include X-rays and electrons. This statement carries the caveat that we must still work with a crystal, and it can be argued that the lattice packing arrangement may force onto a molecule some of its constituent atom's positions and/or dynamics that are not present in *in vivo* conditions. In our crystallographic databases neutron crystal structures are by far the smallest in number due to the practical constraints of low neutron beam fluxes and long measuring times as well as fewer instruments available globally. Not giving a proper attention to the truth of neutron crystal structures is then 'not fair', to use the social scientists' term.

Likewise, we can see the relevance of van der Aalst *et al.*'s (2017) conception of accuracy in our practices. They stress the need for:

> *"not just presenting results or make predictions, but also explicitly [providing] meta-information on the accuracy of the output."*

In this context van der Aalst et al's (2017) perception of trust is also very appropriate:

> *"The journey from raw data to meaningful inferences involves multiple steps and actors, thus accountability and comprehensibility are essential for transparency."*

The translation of this into the crystallographic sciences is perhaps best illustrated by considering structural dynamics, where changes of structure under a perturbation (such as using light in photo-crystallography) are small. Repeat processing of raw diffraction data using different softwares might be selected to find a structural change. Therefore, the availability of raw diffraction data allows a full comparison of different softwares' results, thereby establishing an estimation of the variance of atomic positions and/or B factors determined from those different raw diffraction data processing workflows.

The least generally applicable FACT criterion is that of confidentiality, which is most relevant to human behavioural or medical information. Nevertheless, related properties such as respecting intellectual property rights or providing access control to restricted subsets within an Open Data ecosystem (*e.g.* data sets held in a repository under a pre-publication embargo) should also be characterised by appropriate metadata within an Open Data management framework. In this ecosystem it is summed up in the maxim that data be as open as possible and as closed as necessary (see below).

4. Raw data measuring hardware: sources and detectors

Let us turn now to considering the practicalities within current and planned crystallographic research areas.

In the past decade or more major changes have been made in both source and detector capabilities. Extremely bright sources of synchrotron radiation have emerged led by MAX IV and ESRF (renamed the ESRF EBS) and similar upgrades are being applied at synchrotron facilities worldwide; the latest gains in source brightness are factors of 100 and across a wide range of photon energies. These gains are in addition to the previous frontier (low emittance) performance of PETRA III available for the past 15 years or so. But PETRA III is now also to be upgraded into an even higher brightness PETRA IV, which will then leapfrog the 'extremely bright sources' yet again, promising high brilliance as well as increased flux, in particular at higher energies.

One theme for many decades in macromolecular crystallography has been to collect 'ideal data' using high photon energies (Helliwell *et al*., 1993, Storm *et al*., 2021). By exploiting the more favourable ratio of elastic scattering to photoelectric absorption at higher energies as well as detectors with high $Z$ sensor materials, data sets of higher quality can be collected (Dickerson & Garman, 2019). Nevertheless, exposure times in high-energy experiments currently need to be significantly longer than for standard energies. This is due to two main reasons: firstly, most beamlines are not currently optimized for high energies, resulting in a significantly lower flux at higher energies. Secondly, the diffraction cross-section decreases with increasing energy, requiring a higher input of photons to get the same amount of diffracted photons. At the moment, there are only very few beamlines which deliver a significant photon flux at energies higher than 20 keV and which are equipped with suitable detectors.

In another theme, smaller crystal sample volumes can be measured with a brighter X-ray beam. However, microcrystals of biological macromolecules are already viable (for a review see Evans *et al.,* 2011) and can be investigated using a specialised beamline such as VMXm (Crawshaw *et al*., 2023). In any case other experimental modalities have come to the fore such as using X-ray lasers or in electron crystallography, both of which yield diffraction data from sub-micron to nanometre sized crystals. So, a likely application of the extremely bright sources is getting protein structures from extremely small crystals exploiting photoelectronic escape (Storm *et al*., 2020) and pushing the time resolution of dynamic crystallography to shorter time intervals. This means higher data rates provided the detectors can handle such rates; and indeed, they are currently managing to do so, as illustrated in Fig. 1.

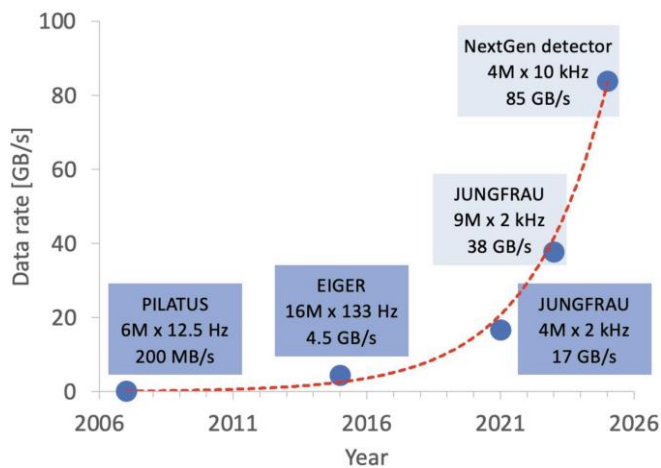

Figure 1. Data rates in Gb/s versus calendar year of successive generations of pixel area detectors. Reproduced with the permission of Filip Leonarski (Leonarski et al., 2023).

However, can tape storage cope with such enhanced data rates? There are substantial increases in tape storage capacities planned; see section 5. Overall, however, these enhanced data rates will challenge the current facility data archiving policies; see section 8.

5. Raw data archive hardware

Since the early investigations into the topic of raw data archiving by the DDDWG (Terwilliger, 2014; Guss & McMahon, 2014), data capture rates and data volumes have increased dramatically, especially at large-scale facilities, and more attention is being given to the economic constraints on long-term archiving, as well as to the environmental impact of maintaining large server or tape storage inventories.

Prompted in part by the recommendations of the DDDWG (2017), institutional archives are reviewing their criteria for long-term retention. There is general recognition that raw diffraction data sets that are associated with a publication merit longer-term retention in order to satisfy the FAIR principles that support validation and reuse (Wilkinson et al., 2016).

The DDDWG recommended also that raw diffraction data sets for currently unsolved crystal structures, or that showed significant diffuse scattering, should also be archived. The recently launched *Raw Data Letters* section of the journal *IUCrData* (Kroon-Batenburg *et al.*, 2022) provides a vehicle for descriptive articles that should identify such data sets, and which helpfully (from the viewpoint of identifying them as candidates for long-term archiving) links them to a peer-reviewed publication.

We consider the physical repositories currently available to users in three categories: (i) general-purpose or domain repositories where users may deposit their own data sets, either because these were collected at a home laboratory or because the facility where the experiment was run does not have a satisfactory data retention policy; (ii) institutional or national repositories which collect researcher outputs voluntarily or under mandate; and (iii) the archiving systems at individual facilities.

5.1 User deposition in public repositories

**5.1.1 General-purpose repositories**

Kroon-Batenburg (2019) conducted a valuable survey of raw diffraction image data sets discoverable through the OpenAire and DataCite portals. She identified a number of open-access general-purpose repositories containing such data sets, as listed in Table 1.

Table 1. Open-access general-purpose repositories containing raw diffraction image data sets (after Kroon-Batenburg, 2019; additional information from re3data.org, fairsharing.org and Stall *et al.*, 2023).

| Repository | Description | Funding model | Fees/costs | Size limits |
|---|---|---|---|---|
| Zenodo | Commissioned by the European Commission (EC) and hosted by CERN. Hosts all types of research artefacts and | EC OpenAire Projects; CERN; US National Institutes of Health; Arcadia Fund; Alfred P. Sloan Foundation; | Free of charge | 50 GB per record (higher quotas can be requested on a discretionary basis) |

|  | accepts all file formats. | donations via CERN & Society Foundation |  |  |
| --- | --- | --- | --- | --- |
| Figshare | Owned by Digital Science, a subsidiary of Springer Nature. Accepts data, papers, code, media and other research outputs. | Commercial; provides research infrastructure services to institutions and publishers. | Free of charge for small data sets (<20GB) | 20 GB |
| Figshare+ | Owned by Digital Science. | Data publishing charge from depositors. | USD 240 for up to 100 GB, then USD 875 per 250 GB | 20 GB – 10 TB |
| Dryad | Non-profit membership organization. Provides a curated general-purpose repository of research data underlying scientific and medical publications. | Costs covered by institutional, publisher and funder members, otherwise a data publication charge | One-time fee of USD 150 for authors | 1 TB per data set |
| Mendeley Data | Owned by Elsevier as part of | Subscription model for | Free of charge | 10 GB per data set |

| | Digital Commons repository family. | academic and government entities | | |
|---|---|---|---|---|

There is clear evidence that the number of diffraction image data sets deposited in these repositories is increasing, but it is difficult to quantify the number currently held, because the different repositories do not offer a suitable search filter. There do not appear to be metadata fields that allow specification of the nature of the study with which a deposited data set is associated.

The actual hardware stack used by these repositories is not easily found. Zenodo resides physically in the CERN Data Centre, currently using an 18 PB disk cluster. While the CERN primary storage infrastructure currently totals 150 PB of data with an expected growth of 30–50 PB per year, it is posited that Zenodo might in future move some or all of its content to offline tape storage (see https://about.zenodo.org/infrastructure). Figshare runs on Amazon Web Services (https://help.figshare.com/article/figshares-approach-to-security-and-stability). Dryad data are hosted on the California Digital Library multi-campus Merritt Repository (https://datadryad.org/stash/mission). Mendeley Data also runs on Amazon Web Services (AWS) but is additionally archived with the Data Archiving and Network Services based in the Netherlands. While additional capacity on cloud-based services can usually be purchased and added easily to an existing service, it is possible that migration of content to offline tape storage might become necessary on economic grounds beyond a certain point (as already noted in the case of the locally-hosted Zenodo). In the case of AWS, tape retrieval from the associated Amazon Glacier storage service may take several hours. As use of these platforms grows across different domains, their policies may change to favour disciplines or types of data set that place greatest demand on their services – in other words, if crystallography supplies a small proportion of their content (whether judged by storage volume, number of distinct data sets or other criteria), it may have little say in the evolution of a platform as a whole.

**5.1.2 Domain-specific repositories in biological crystallography**

There are currently four active repositories for raw diffraction data from macromolecular crystallography experiments: the SBGrid Data Bank, the Integrated Resource for Reproducibility in Macromolecular Crystallography (IRRMC), the Xtal Raw Data Archive (XRDa) hosted by the Japanese partner of the Worldwide Protein Data Bank, and the Macromolecular Xtallography Raw Data Repository (Table 2).

Table 2. Open-access raw diffraction data repositories for biological crystallography.

| Repository | Description | Funding model | Number of data sets |
| --- | --- | --- | --- |
| SBGrid Data Bank | Community-driven repository for X-ray diffraction (also microED and lattice light-sheet microscopy) data sets in structural biology. | Subscription model for SBGrid Consortium members. | 819 |
| IRRMC | Hosted by the Minor Laboratory of University of Virginia; open to any data submissions related to structures deposited with the Protein Data Bank. | Targeted Software Development award as part of the BD2K (Big Data to Knowledge) program of the National Institutes of Health. | 9648 |
| XRDa | Aims to collect raw crystal diffraction data for entries submitted to the Protein Data Bank, as well as independent | Supported by National Bioscience Database Center, Japan Science and Technology Agency, and by public donations. | 114 |

| | submissions. Hosted by PDBj. | | |
|---|---|---|---|
| MX-RDR | Archives and provides access to raw diffraction data collected for macromolecular crystals. Includes tools for creating datasets of crystallographic metadata by combining information extracted directly from diffraction images and obtained from a PDB deposit and/or user input. | Developed as a part of an EU funded project, coordinated by the Interdisciplinary Centre for Mathematical and Computational Modelling, University of Warsaw. | 417 |

These archives are better suited to matching the evolving requirements of crystallographic research, and in particular their funding models might be expected to respond to the perceived needs of the active community. However, none operates on a strictly commercial basis, and so all are vulnerable ultimately to changes in public funding policy.

5.2 User deposition in private repositories

We use the term 'private' to refer to institutional or national repositories designed to store and monitor research outputs from academic staff. We do not mean to imply that the data are not made public; many institutions provide open access to deposited datasets to honour the FAIR principles such as the excellent work of the University of Manchester in this respect (Kroon-Batenburg *et al.*, 2017).

There is significant diversity in the policies and capacities of such institutions, and so we cannot draw general conclusions about their significance. Some may host copies of data sets that have also been stored in public archives. On the one hand, this increases resilience through redundancy of information; on the other hand, it complicates maintenance and raises the prospect of diverging versions.

Although these facilities increase the number of possible storage repositories, they also suffer from the shortcoming already identified, namely the difficulty in discoverability of data sets associated with specific types of research output. Many such institutions, for instance the ESRF, do issue DOIs or other persistent identifiers for deposited material, so that links from the published literature do establish one 'findable' route. However, it is still not possible to browse or interrogate any individual repository to retrieve only data sets of a specific type (especially if they are not associated with published articles), although a cross-repository search interface is available through DataCite Commons (https://commons.datacite.org). Also the archive mining tools seem now to be available (see e.g. https://www.ch.cam.ac.uk/person/pm286 ) and text mining is ongoing in earnest (see e.g. http://core.ac.uk) as is data mining (see e.g. http://chemdataextractor.org/docs/intro ).

Many of the repositories do support established protocols such as the Open Archives Initiative Protocol for Metadata Harvesting (OAI-PMH) (https://www.openarchives.org/pmh/), but we are not yet aware of any concerted efforts to introduce more granular discoverability through use of existing features within this protocol, such as the extension to the existing 'set' construct, as suggested by Guss & McMahon (2014).

5.3 Data archiving at experimental facilities

While users of synchrotrons and other large experimental facilities will usually take copies of their collected data sets back to their home institutions, there is pressure on the facilities to offer archiving services, partly for the benefit of users who lose their own copies, or in some cases to facilitate analysis of the data with in-facility software and computing resources; but increasingly to provide repositories from which raw data may be accessed under the FAIR principles.

While facilities are generally equipped with high-performance computing platforms designed to handle the ever-growing data transfer rates of each new generation of detector, the prospect of storing large volumes of collected data for long periods is becoming increasingly challenging.

Several of the presentations at the Melbourne Workshop illustrated the scale of the challenge. For example (Barty, 2023), the PETRA-III source at the German synchrotron currently stores ~4.5 PB on disk-based (GPFS filesystem) storage in the course of a year, with a 180-day retention policy, then writes a dual tape copy (2 x 6 PB per year). The projected upgrade to PETRA-IV is anticipated to generate enough data by 2028 to require over 500 PB of disk storage and up to 1 exabyte of tape storage per year if the same data retention approach is maintained. The power requirements for data storage alone are projected to exceed 1 MW.

Fig. 2 demonstrates the cascade of stored data to slower but higher-capacity systems at the European X-ray Free-electron source at Schenefeld, which uses the DESY data centre storage infrastructure. It is notable that this model is common to many facilities, but practice is variable; for instance, at the Paul Scherrer Institute only single tape copies are retained to reduce costs (Ashton, 2023). As pressure grows to reduce the energy footprint of large-scale facilities (Abela *et al.*, 2023), there is little doubt that more data storage will be transferred to magnetic tape. However, apart from increasing the time needed to access data stored on tape, media costs are also substantial – the PETRA-IV case study cited above projects renewables costs of EUR 50 million per year by 2028.

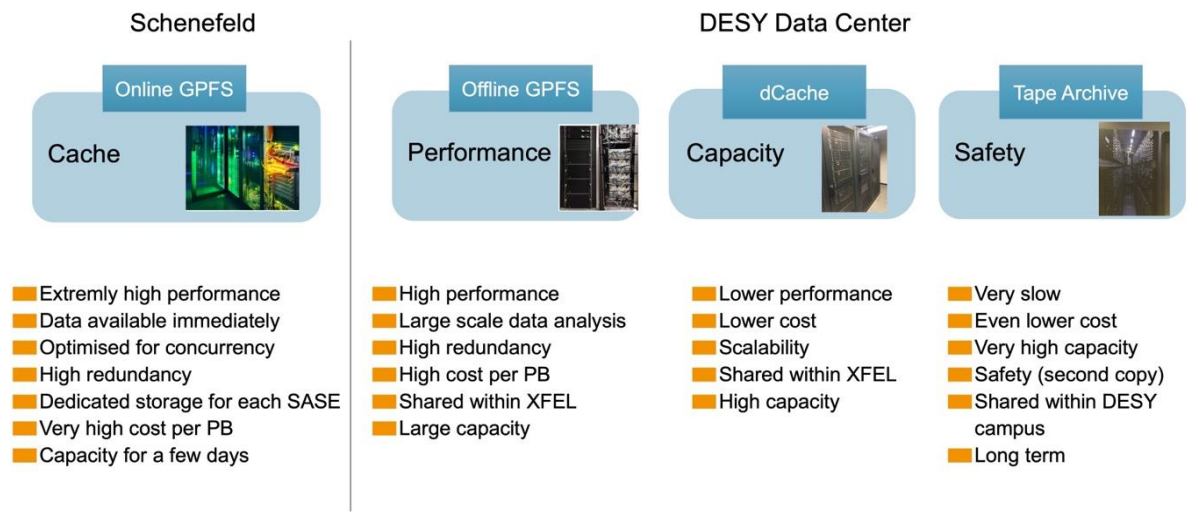

Figure 2. Example of a cascade of data storage approaches. Experimental data are captured to very high-performance systems in real time, but move progressively to slower but higher-capacity media as the focus shifts from processing and analysis to review, backup and potential reuse. From Dall'Antonia (2023), originally created by Krzysztof Wrona, European XFEL.

It therefore seems inevitable that pressure will grow on the facilities to store a smaller proportion of the raw data generated from experiments on a long-term basis. Already facilities are beginning to process raw data from serial crystallography experiments on the fly, and to consider various strategies for retaining progressively smaller quantities of data: store hits only; store indexed frames only; use lossy compression methods; store data only when it yields results; store a random sample of the data (Tolstikova, 2023).

It should also be borne in mind that for many facilities crystallographic experiments do not supply the bulk of the data collected (*e.g.* in imaging experiments), so that there is a danger that the scientific desirables of the crystallographic community might come into competition with the overall economic pressures on the facility.

6. Raw data processing software

Most software for data processing of single crystal data is designed for rotation scans using area detection systems. Two dimensional detectors were developed because of the need for rapid data collection to avoid radiation damage in particular for large unit cells (Arndt&Gilmore, 1979). Processing data from the oscillation (or rotation) method presented specific problems, because of the partiality of reflections as they move through the Ewald sphere (Rossmann, 1979; Arndt & Wonacott, 1977). Post refinement techniques (Rossmann, 1985) that allow the refinement of partiality based on crystal orientations, beam divergence, wave-length dispersion and crystal mosaicity, their relation being detailed in papers by Greenhough and Helliwell (1982a, 1982b), turned out to be very powerful. In addition, Rossmann (1985) introduced profile fitting for quantitative analyses of reflection data. Data processing includes the following steps: peak searching, indexing to find the unit cell, space group determination, determination of Bragg intensities with box summation or profile fitting and scaling to bring all reflection data on a common scale including correction for background, radiation damage and absorption correction. The workflow of crystallographic data collection and processing is different for researchers equipped with in-house diffractometers and those that use synchrotron beam lines.

Diffractometer vendors deliver fully integrated systems with sample management data bases, data collection control software, often for multi-circle goniometers, data storage facilities on local computers and data processing software. The most frequently used in-house systems currently are the Bruker series of diffractometers with *Apex/Proteum* data processing software based on *Saint* (Bruker, 2019), Rigaku Oxford Diffraction diffractometers with *CrysAlisPro* software (RigakuOD, 2019) and Stoe diffractometers with *X-AREA* software (Stoe & Cie, 2016). The equipment and software can be optimized for chemical and macromolecular crystallography.

For synchrotron beamlines, mainly driven by macromolecular crystallography, several software packages were developed: *Mosflm* (Leslie, 1999) mostly used via its graphical interface *iMosflm* and its incorporation in the CCP4 Program Suite (Winn et al., 2011), *HKL2000* (Otwinowski & Minor, 1997) which is installed at many US synchrotron beamlines and *XDS* (Kabsch, 2010) which performs well in unattended data processing and is often used in automatic data processing pipelines. All of these packages use profile learning techniques to accurately integrate the Bragg spots. *XDS* transforms pixel data to reciprocal space and encourage fine-slicing to reduce experimental broadening effects in the spot profiles, making profile learning more robust. *EVAL* (Schreurs et al., 2010) uses a ray-tracing profile simulation technique, is very versatile and capable of treating many complicated diffraction problems but has a steep learning curve. At Diamond Light Source, Lawrence Berkeley National Laboratory and CCP4 a new software suite, *DIALS*, for the analysis of crystallographic X-ray diffraction data was developed (Winter et al., 2018) that is set up in a completely modular way and built upon the *cctbx* library (*Computational Crystallography Toolbox*; Grosse-Kunstleve *et al.*, 2002). Users may choose any of the above packages based on their computer operating system, wish for graphical interaction with data processing or because it is installed at the synchrotron beam line.

Initially, researchers would transfer their collected data frames by slow internet connections or on DVD to their home computer and process them locally. However, in recent years the workflow has changed considerably. Most MX beamlines have developed automatic processing tools that streamline the generation of input parameters and automatically guide the user through the data reduction steps, resulting in a processed diffraction data file in mtz format. There is no need for the user to transfer the raw images to their home computer; the mtz file suffices. This transfers the responsibility for data archiving and curation to the synchrotron facility. Some examples of such pipelines are experiment-control and sample management systems, such as ISPyB (Beteva *et al.*, 2006) now including automatic processing pipelines: GrenADES (Monaco *et al.*, 2013), EDNA (Incardona, 2009), AutoPROC (Vonrhein, 2011) at ESRF and xia2 (Winter *et al.*, 2013) at DLS. SSRL uses Web-Ice (Gonzalez, 2008) for integrated data collection and analysis.

With the advent of extremely bright sources of synchrotron radiation (see section 4) data are collected on ever smaller crystals that easily suffer radiation damage. This is solved by multi-crystal data collection and combining the data, so-called serial synchrotron crystallography (SSX). This technique is inspired by serial femtosecond crystallography with X-ray free electron lasers (SFX), where hundreds of thousands or millions of micro- or even nano-sized crystals are irradiated and diffract before they are destroyed by radiation damage, allowing time-resolved room-temperature data collection. SFX produces a huge number of still diffraction images and new methods and software had to be developed to index the still images and correct for the intrinsic partiality of the Bragg reflections: *CrystFEL*

(White *et al.*, 2012), DIALS (Winter *et al.*, 2018) in combination with *cctbx.xfel* (Hattne *et al.*, 2014) and *nXDS* (Kabsch, 2014). The sheer volume of data produced for a single project and per day of running the facility is enormous, and requires advanced strategies for managing and curating the data (Barty, 2023) (see section 4).

7. Raw data peer review

There is no value in publishing raw data, whether in a traditional journal or through deposition in a public archive, unless the data are reusable. Reusability has two aspects: (i) correct description of the data, and (ii) sufficient information about the experimental conditions and sample from which the data have been collected. The task of the reviewer is to ensure that both of these criteria have been met.

Correct data description is largely a matter of meeting technical requirements. Journals and archives can therefore simplify the work of the raw data reviewer by limiting the range of accepted data standards and providing automated checks: much of the quality guarantee provided by the journal or archive then relies on the quality of the checking software and selection of appropriate data standards, with the reviewer providing the final assessment based on the output of automated tools.

Sample description is less susceptible to automated checks. The sample provenance needs to be sufficiently-well described to repeat the experiment. Any machine-readable standard for sample description therefore needs to cover sample origins as widely varying as complex synthesis and crystallisation procedures, geological field trips, and bespoke industrial manufacturing equipment. Furthermore, while there are machine-readable standards covering particular sample creation procedures, it is not reasonable to automatically reject a paper if such standard information is missing, as the sample may have a novel origin not envisioned by available standards. The simplest option is therefore for the reviewer to manually assess the description of the sample for completeness, a task that is familiar from refereeing conventional papers.

Description of the experimental environment is similarly too open-ended to be covered by currently extant standards. While commonly-varied parameters, such as temperature and pressure, are included in most metadata standards, specialist techniques such as pump-probe have not yet been adequately formalised. Until standards bodies catch up with developments in experimental techniques, adequate description of the sample environment will remain on the human reviewer's checklist.

The review process for single crystal datasets in *Raw Data Letters* addresses the above considerations in the context of a traditional journal. Dataset submissions are restricted to open standards for which required metadata have been defined, in this case files meeting either the NXmx "gold standard" (Bernstein *et al.*, 2020) or equivalent imgCIF standard (Bernstein, 2006). The journal provides an online tool which generates several check images, based on which the correctness of the geometry descriptions and wavelength can be immediately assessed. Reviewers, of course, retain the option of ingesting the dataset into their own software to determine acceptability. *Raw Data Letters* has a further requirement that the dataset have some intrinsically interesting features, which is a judgement best left to human reviewers.

In the context of public archives of raw data, curation performs a similar function to peer review for journals. The IMMRC (see section 5.1.2 above) provides a detailed description of the data curation process (Grabowski et al, 2016), which includes metadata harvesting from a variety of sources, followed by a series of checks. As IMMRC contains only structural biology datasets, which are generally linked to wwPDB depositions containing detailed sample provenance information, there is no need for the raw data deposition to include sample- or environment- related metadata.

Post-publication data review is also a viable route for raw data publications. For example, the SBGrid data archive (Meyer et al, 2016) publishes datasets from registered users without assessing data reproducibility, instead reporting on data reprocessing outcomes post-publication. As for IMMRC, links to wwPDB depositions ensure that adequate descriptions of the sample have been provided.

8. The evolution of data policy at photon and neutron central facilities

Central facilities providing X-rays and neutrons used in crystallography, diffraction and scattering today occupy a major role. The extent of this role varies a lot between the sub-disciplines of these fields. For example, for macromolecular crystallography around 90% of all depositions into the PDB are from synchrotron facilities. For chemical crystallography the Cambridge Crystallographic Data Centre has approximately 97% from home laboratory X-ray source measurements. This simply reflects the much larger scattering strength of smaller unit cells seen in chemical crystallography, simpler means of crystallographic structure-factor phase determination, and thus experimental requirements that do not usually include high-intensity tuneable beams at a synchrotron beamline.

The evolution of photon and neutron central facilities' data policies in the past two decades or so has seen substantial changes. These changes have reflected the practicalities as highlighted in other sections of this article. However, there is also, first of all, an evolution of policy thinking especially by the funding agencies as they increasingly realised that some commercial publishers were making large profits out of taxpayers' funded research. Clearly this was a violation of principle not least that a member of that tax paying public could not access the research results in scientific journals that the funding agencies as their proxies had funded. An awkward point in this simple and obviously compelling argument was that the funding agencies typically funded 'only' about 20% or so of the proposals put to them. So, what about the results from unfunded research? A second aspect was that many of the learned societies had their own journals which made only small surpluses which in any case was invested in schemes like student bursaries for their training. Nevertheless, an unstoppable momentum has built up in ensuring open access to all research results. That these results should be presented along with their underpinning data has been a tradition of crystallographers, introduced by Bragg (1913) and formalised in the crystallographic databases firstly with the Cambridge Structural Database launched in 1965 and the PDB launched in 1971. A wide spectrum of databases is available today as summarised in Bruno *et al*. (2017). A landmark in policy development was the 2007 report of the OECD, already mentioned in section 1. Significant advances in policy development followed the publication of this report that sought to improve the practice of global science through recommendations on access to publicly-funded research data. Their focus was on computer-readable data that were the primary sources for scientific research and thus appropriate for validating research findings.

In the final section on Sustainability, the report states that:

> *"[sustainability] can be a difficult task, given that most research projects, and the public funding provided, have a limited duration, whereas ensuring access to the data produced is a long-term undertaking."*

These guidelines from the OECD presented a challenge, as well as opportunities, to all universities, principal investigators, and the central facilities. It also challenged the funding agencies who had to give a clear (or clearer) budget line to data management, storage, and access costs. A common theme emerged, in mainland Europe at least, consistent with the OECD (2007) guiding principles, that: -

> *"measured data will be retained by a facility for at least ten years"*
> and
> *"measured data will be made public after three years"*.

9. Possible future developments: Data flows outstrip data storage capacities

With ever increasing data rates and more and more insecure funding perspectives, there might come a point in time where data flows outstrip data storage capacities. This might in particular be true for serial crystallography experiments. As those experiments produce the most data in crystallography by far, new procedures could significantly reduce the amount of data to be stored, as outlined above. A lot can also be gained by storing the data in a compressed format. For standard crystallography, the volume of data to be stored is relatively modest compared with other disciplines. Should one still accept the need to reduce their volume, some prerequisites should be met when realizing a more economic form for storing the results from experiments. One of them is the availability of electronic logbooks to enable full transparency on the protein's crystallisation conditions. Combining this with the realization of the gold standard for metadata (Bernstein *et al.*, 2020), it might be sufficient to only document the software applied and save the structure factors as a result of the experiment. This, of course, supposes both that accessible software archives are maintained for the long term, and that specific version snapshots can be retrieved to match the original processing workflow. However, this prevents reprocessing of data with software yet to be developed. Helliwell (2023) has discussed the challenges for quantifying small structural displacements, and their error estimates, that can be the situation for structural dynamics studies.

10. Conclusions

Despite all the progress in instrumentation, detector technology and software, raw data still represent the ground truth. In addition, the often overlooked or under-processed raw data harbour hidden treasures, unlocking potential insights that might have been missed in the initial analysis. In MX, for example, a new step forward taken recently is making advanced exploitation of processed but unmerged reflection intensity data during processing and then model refinement (Vonrhein et al 2024 Acta D in press). Crucially, raw data serve as a fundamental tool for training and education in the field. Providing aspiring researchers with access to the unfiltered intricacies of crystallographic experiments nurtures a deeper understanding and proficiency in the methodology.

In the realm of software development, the untapped potential within raw data emerges as a catalyst for innovation. The data may contain hidden patterns or information yet to be extracted, pushing the boundaries of what current analytical tools can reveal. To cover all these aspects appropriately, *Raw Data Letters* was founded recently.

Raw data also play a pivotal role in safeguarding against fraud. By maintaining transparency and authenticity in the data collection process, the scientific community fortifies itself against misleading or fabricated results. This becomes even more crucial in the age of artificial intelligence (AI). On the other hand, AI and machine learning in particular, offers new opportunities in the domain of raw data mining as well as text and processed data mining. In combination with electronic logbooks the capabilities of AI could be enhanced and contribute to the reproducibility of high-quality data. This synergy propels scientific advancements and reinforces the reliability of crystallographic research.

As the activities of the DDDWG and CommDat in the past decade have demonstrated, the relevant cost–benefit analyses for archiving raw diffraction data are complex and must constantly take account of changing technologies and practices. They are also subject to available funding, which is not always under the control of the scientific community.

Nevertheless, our continuing efforts to update such analyses will be important in informing public funding policies.

As the scientific landscape evolves, the discussion surrounding what to store in the context of such as serial crystallography becomes paramount. Continued and intensified deliberations on this front are essential for adapting to new methodologies and ensuring the seamless progression of crystallographic research.

**Acknowledgements**

We thank all colleagues on the IUCr DDDWG and the Committee on Data for joint work and discussions as well as the participants at all of their workshops. We appreciate the work of the Melbourne IUCr 2023 Congress Organisers in producing such an excellent discussion framework.